\documentclass{emulateapj}

\shorttitle{Erosion driven size redistribution of solids}
\shortauthors{Grishin, Rozner and Perets}

\usepackage[breaklinks,colorlinks,citecolor=blue]{hyperref}
\usepackage{graphicx}	
\usepackage{amsmath}	
\usepackage{amssymb}	
\usepackage{color}
\usepackage{float}

\begin{document}


\title{Erosion driven size-redistribution of protoplanetary disk solids\\ and the onset of streaming Instability and Pebble Accretion}


\author{Evgeni Grishin, Mor Rozner and Hagai B. Perets}
\affil{Technion, Israel Institute of Technology, Haifa, Israel 3200003; eugeneg@campus.technion.ac.il}

\begin{abstract}
The formation of the first planetesimals and the final growth of planetary cores relies on the abundance of small pebbles. The efficiencies of both the streaming instability (SI) process, suggested to catalyze the early growth of planetesimals, and the pebble-accretion process, suggested to accelerate the growth of planetary cores, depend on the sizes of solids residing in the disk. In particular, these processes were found to be sensitive to size-distribution of solids, and efficient planetesimal formation and growth through these channels require a limited pebble size-distribution. Here we show that aeolian-erosion, a process that efficiently grinds down boulders into a mono-sized distribution of pebbles, provides a natural upper limit for the maximal pebble sizes (in terms of their Stokes number).  We find the dependence of this upper limit on the radial separation, disk age, turbulence strength, and the grain-size composition of the boulders in the disk. SI is favourable in areas with Stokes number less than 0.1, which is found in the inner sub-au regions of the disk. This upper limit shapes the size distribution of small pebbles and thereby catalyzes the early onset of planetesimal formation due to SI, and the later core accretion growth through pebble accretion. 

\end{abstract}

\keywords{planets and satellites: formation --- protoplanetary disks --- planet–disk interactions}


\section{Introduction}

The early stages of planet formation occur in protoplanetary disks around young stars, which initially contain mostly gas, and roughly  $1\%$ of dust. Planet formation takes place over many orders of magnitude, beginning with micron sized dust grains which collisionally grow to cm-sized pebbles, and later grow into km-sized planetesimals, and eventually  form planetary embryos and planets  \citep{ChiangYoudin2010}. 

Although the early growth of dust grains can be understood through collisional processes, the formation of the first planetesimals proves to be a major challenge. Small grains are tightly coupled to the gas flow and can efficiently grow to mm-cm pebbles. The larger meter-sized boulders are partially decoupled from the gas flow and experience various growth barriers \citep[][and references therein]{bw2000}. In particular, the radial-drift barrier prevents particles from growing beyond cm-m scales, since such boulders are effectively lost to the main star \citep{adachi76,Weidenschilling1977}, and collisional fragmentation limits rapid growth of $\sim$ meter-size boulders \citep{bw2000}. Interstellar planetesimal seeding \citep{EvgeniHagaiSeeding} could provide large enough planetesimals to young systems, thus liberating it from its initial growth barriers. The generation of the first planetesimals, however, is still debated. 

Recently, we had suggested that aeolian-erosion gives rise to an additional potential growth barrier for pebbles/boulders/rocks growth, where beyond a certain threshold velocity, the headwind from the gas flow erodes material from the surface of the boulder, as it overcomes the cohesive forces holding its material together \citep{AeolianErosion1}. The erosion can either grind down larger boulders into smaller pebbles, or set an additional growth barrier for the growing pebbles, even if the other barriers are circumvented.

The streaming instability (SI) \citep{YoudinGoodman2005} is a potentially promising mechanism to overcome the radial drift (and other barriers) to form planetesimals. SI catalyzes the localized concentration of solids in the disk to the point where  gravitational collapse can operate and directly form large planetesimals. Possible observations and simulations that support this channel rely on studies of asteroids size distributions \citep{morbi2009, li2019} and binary Kuiper-belt object binary masses, compositions \citep{nesvorny2010} and orientations \citep{kbospin}. However, the robustness of SI is debated. In particular, SI which leads to produce strong clumping ans successful planetesimal formation requires large metallicity of the protoplanetary disk, local dust to gas ratio above unity and an optimal size of the pebbles and pressure gradients \citep{johansen2009, yang17, so2018}.
 
Early SI studies assumed simple mono-size distribution of solids in the disk. However, recently, \cite{krapp2019} had shown that SI proves to be far-less efficient when multi-size solids distribution is considered.  They find that for sufficiently wide distribution of pebble sizes, the timescale for the growth of the SI unstable mode is linearly decreasing with the number of species, and does not converge (see Figures 2 and 4 of \cite{krapp2019}). Interferometric and scattered light observations of young disks suggest coexistence of both small $\mu$-sized grain and $\sim$ cm sized pebbles \citep{menu14,vb17}. Thus, the existence of a wide size-distribution, typically expected in planet formation models \citep{bai10,sch18} could severely limit the applicability of the SI scenario.

At later stages, the formation of gas/ice giants requires the growth of planetary cores in the standard core accretion scenario \citep{pollack96}. The source of the accreted solids was first attributed to planetesimals, but the accretion rate was found to be too slow as to efficiently grow planetary cores at large separations. However, it was later suggested that wind-assisted accretion of pebbles could provide a more efficient channel for planetary accretion and growth 
\citep{ok2010,PeretsMurray2011,lj2012}. The growth rate and hence the final embryo/planet mass depend on many parameters, including the pebble sizes and abundance, the location in the disk, core formation times \citep{bitsch2015,vo16, ol18}, turbulence levels \citep{RMC18,RMC19} and planetary envelope structure and evolution \citep{ll17, vazan18}.

Both the early planetesimal formation via SI and later subsequent formation of planets due to core accretion rely on the flow of pebbles. Only pebbles of a certain size range, pending on disk model and radial location, can significantly contribute. Thus, concentration of pebbles of similar sizes in a localized region in the disk could be beneficial for the formation and growth of planetesimals/planets \citep{loj19}.

Various mechanisms for concentration of particles have been suggested, including vortices \citep{barge95, rkl2015}, zonal flows \citep{youdin09}, pressure bumps \citep{pinilla12,zhu12} or planetary torques \citep{bp18,cl18}. These mechanisms involve either complex turbulent magnetohydrodynamical effects and/or pre-existing planets and studied mostly numerically. Here we present a simple, analytic model for the redistribution of disk solids sizes due to a different mechanism, namely  aeolian-erosion. 

In this letter we utilize the aeolian erosion barrier as a natural source of size-segregation and concentration. We focus on the first stages of planet formation assuming no planets or pressure bumps are present. We consider laminar disks flow, and later discuss turbulent disks. In sec. \ref{s2} we review the aeolian-erosion mechanism and derive the upper limit for the critical Stokes number of surviving solids as a function of the radial location on the disk and the size of the detached grains (i.e. assuming pebbles/boulders are composed of small grains of some typical size, which are removed by the head winds) for laminar and turbulent flows. This, in turn, effectively determines the maximal size of eroded pebbles that survive in the disk. We discuss the implications of the aeolian-erosion pebble size-limit for the SI and pebble accretion processes in sec. \ref{s3} and summarize in sec. \ref{s4}.

\section{Critical Stokes number from Aeolian-Erosion} \label{s2}

\subsection{Aeolian Erosion}
In \cite{AeolianErosion1}, we introduced and discussed the concept of the aeolian-erosion barrier. As small pebbles grow into boulders they are held by cohesive forces. The wind from the gas flow can detach dust-grains and pebbles from the surface of the growing boulder. The threshold relative wind velocity at the point when the shear pressure overcomes the cohesion and detach particle from the boulder surface is derived from \cite{sl00} 
\begin{equation}
v_{\rm th} = \sqrt{\frac{A_N \gamma}{\rho_g d}}, \label{eq:vth}
\end{equation}
where $\rho_g$ is the local gas density and $d$ is the typical size of the grains composing the pebble. $A_N$ is a dimensionless number that depends on the Reynolds number, and $\gamma$ is the surface energy. Wind tunnel experiments found a good fit with a constant value of  $A_N = 1.23\cdot 10^{-2}$ and $\gamma$ in the range of $1.65-5 \cdot 10^{-1} \ \rm g\  s^{-2}$ for grain sizes in the range of $50 -1800\ \rm \mu m$, \citep{iversen82}. Recent microgravity experiments of silicate glass spheres measured the surface energy in the range of $\gamma = 7.8\pm3.8 \cdot 10^{-2} \rm \ g\ s^{-2}$ \citep{demirci20}. We choose $\gamma=1.65 \cdot 10^{-1} \rm \ g\ s^{-2}$ to be compatible with both experiments.

When the relative velocity exceeds the threshold velocity, grains from the outer layer of the pebble/boulder are removed and the mass loss rate is fast. The erosion timescale is   \citep{AeolianErosion1}. 
\begin{align}\label{eq:char_timescales}
 t_{\rm ero}= \frac{R}{|\dot R|} = \frac{4\pi R^2 \rho_p F_{\rm coh}}{\rho_g v_{\rm rel}^3 m_d},
\end{align}
where $R$ is the size of the body, $v_{\rm rel}$ is the relative velocity, $F_{\rm coh}$ is the strength of the cohesive forces and $m_d$ is the mass of the released grains. The cohesive force scales as $F_{\rm coh} \propto d$, the grains size, with a proportionality constant around $10^2 \rm g\ s^{-2}$, determined from experiments (see \citealp{AeolianErosion1} and \citealp{sl00} for details and references.) Generally, the erosion will be very fast, comparable to dynamical timescales for particles less than $\lesssim 10\ \rm m$ (see e.g. Fig. 2 of \citealp{AeolianErosion1}), which is comparable to the rapid erosion rates determined in wind tunnel experiments of \cite{paraskov2006}, and more recent microgravity experiments of \cite{demirci20}. The mass loss continues until  the relative wind velocity (which changes due to the continuous decrease in the size of the eroding pebble) becomes smaller than the threshold velocity.

The gas flows in a sub-Keplerian velocity determined by the pressure gradient profile and the location in the disk. The deviation from Keplerian velocity is measured by $\eta \propto (h/a)^2$, where $h$ is the scale height and $a$ is the distance from the star. Using polar coordinates, the components of relative velocity between the object and gas are (we generally follow the same disk model as assumed in \cite{PeretsMurray2011} and references therein)

\begin{align}
    v_{\rm rel,r}=-\frac{2\eta v_k \tau_s}{1+\tau_s^2}, \ v_{\rm rel,\phi}= -\eta v_k \left(\frac{1}{1+\tau_s^2}-1\right), \label{eq:vrel}
\end{align}
where the Stokes number is defined by

\begin{align}
    {\tau_s}= \Omega t_{\rm stop}; \  t_{\rm stop}= \frac{mv_{\rm rel}}{F_{D}}, \label{eq:st}
\end{align}
where $\Omega$ is the angular Keplerian velocity, $v_k$ is the Keplerian velocity, $m$ is the object's mass and  $t_{\rm stop}$ is the stopping time. $F_D$ is the aerodynamic drag force.  

\begin{figure*}
  \includegraphics[width=1.\linewidth, width=8.8cm]{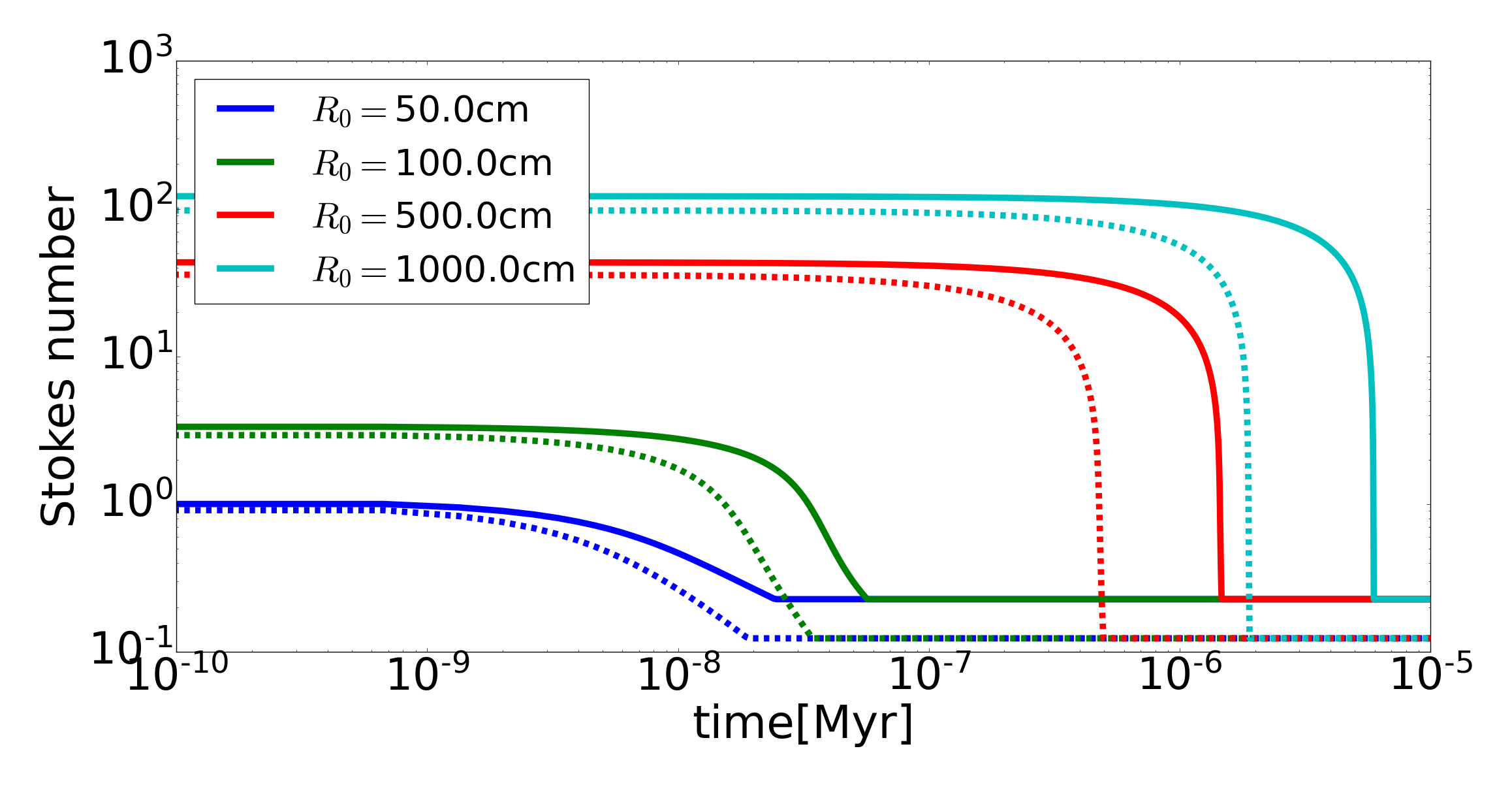} \includegraphics[width=1.\linewidth, width=8.8cm]{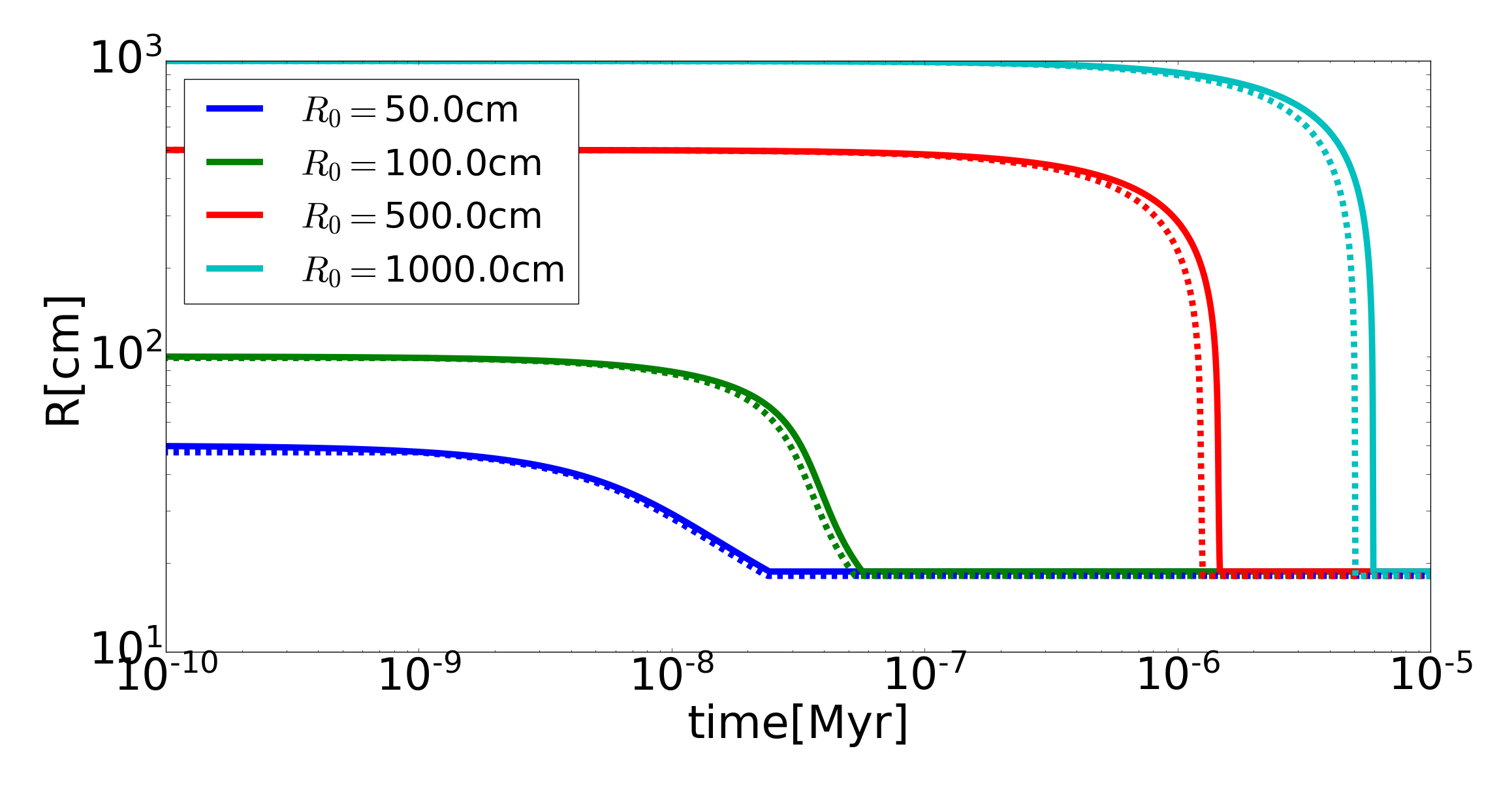}
\caption{The time-evolution of the Stokes number (left) and of the body size (right) on the initial size of the body at a fixed distance of $a=1\ \rm{au}$ from the star, and dust grains of size $d=0.1\ \rm cm$. Solid lines correspond to integration with the laminar relative velocity. Dashed lines depict integration of both laminar and turbulent velocity. } 
\label{fig:dynamical_Stokes}
\end{figure*}

In Fig. \ref{fig:dynamical_Stokes} we show the aeolian-erosion time evolution of bodies of various initial sizes, but using the Stokes number as a measure. In obtaining Fig. 1 we used the flared Chiang-Goldreich disk model (\citealp{gc1997}, see also \citealp{PeretsMurray2011} and \citealp{gp15}), with $\eta\approx 2\cdot 10^{-3} (a/\rm au)^{4/7}$ and $\rho_g = 3\cdot 10^{-9} (a/\rm au)^{-16/7}\ \rm g\  cm^2 $. We used $a=1\ \rm au$ and $d=0.1\ \rm cm$, similarly to our default assumption in \cite{AeolianErosion1}. The final Stokes number will lower for the case of turbulent velocities, as explained below.  

\subsection{Critical Stokes number}
The conclusion from Fig. \ref{fig:dynamical_Stokes} is that the size distribution of particles is limited to a critical Stokes number, $\tau_{\star}$, which depends on the properties of the composing-grains, the sizes of the eroding pebbles and the properties of the disk. For only laminar relatively velocities, we present an analytic solution for $\tau_\star$ as a function of the grain and disk properties. For the turbulent case, we arrive at a fifth order polynomial and find its roots numerically. We discuss the implications for SI later in sec. 3.

\subsubsection{Laminar Case}
The scalar relative velocity from Eq. \ref{eq:vrel} is 
\begin{equation}
v_{{\rm rel}}=\sqrt{v_{r}^{2}+v_{\phi}^{2}}=\eta v_{K} \frac{\tau_s\sqrt{4+\tau_s^{2}}}{1+\tau_s^{2}}\equiv \eta v_k g(\tau_s).
\end{equation}

The erosion is quenched once $v_{{\rm rel}}(\tau_s)\le v_{{\rm th}}$, which defines the critical Stokes number $\tau_{\star}$ as a function of the radial location on the disk. 
By setting the dimensionless laminar relative velocity $\kappa_{\rm l} \equiv \eta v_{k}/v_{{\rm th}}$ the condition  becomes $g(\tau_{\star})=\kappa_{\rm l}^{-1}$. Inverting the equation leads to a second degree polynomial, solved via the standard quadratic formula to yield 
\begin{equation}
\tau_{\star}=\left[ \frac{1-2\kappa_{\rm l}^{2} + \kappa_{\rm l} \sqrt{4\kappa_{\rm l}^{2} - 3}}{\kappa_{\rm l}^{2}-1} \right]^{1/2} \label{tau_crit}
\end{equation}
The existence of a real solution requires $\kappa_{\rm l} \ge \sqrt3/2$. Note that the case $a=1\ \rm au$ and $d=0.1\ \rm cm$ leads to $\tau_\star \approx 0.22$, which is the critical Stokes number in our example in Fig. \ref{fig:dynamical_Stokes}.  

\subsubsection{Turbulent Case}
The disk could also be turbulent. The strength of the turbulence is parametrized by the standard Shakura-Sunyaev parameter $\alpha$. The relative turbulent velocity depends on $\alpha$, and on the dimensionless Stokes and turbulent Reynolds numbers. The turbulent Reynolds number is the ratio of the turbulent to molecular viscosity, or the ratio of the largest eddy to the mean free path \citep{rp18}. In any case, the turbulent Reynolds number is of the order $\sim \alpha \times 10^{10}$ and much larger than any typical Stokes number.

In the limit of infinite turbulent Reynolds number, the turbulent velocity component is given by $v_{\rm turb} = \sqrt{\alpha} c_s \sqrt{\tau_s/(1+\tau_s)}$, where the sound speed is $c_s \approx 6.6 \cdot 10^4 (a/\rm{au})^{-3/14}\ \rm cm\ s^{-1}$. Note that the ratio $c_s/v_k \approx 0.022(a/\rm{au})^{2/7}$ is the aspect ratio of the disk, as set from the disk profile. The total relative velocity is the sum of the squares of the laminar and turbulent velocities, $v_{\rm tot}^2 = v_{\rm rel}^2 + v_{\rm turb}^2$. 

The erosion stops once $v_{\rm tot} \le v_{\rm th}$. Similarly to the laminar case, we can define the dimensionless turbulent velocity $\kappa_{\rm turb} \equiv \sqrt{\alpha} c_s / v_{\rm th}$, and the condition for the critical Stokes number becomes:

\begin{equation}
\kappa_{\rm l}^2 \frac{\tau_\star^2(4+\tau_\star^2)}{(1+\tau_\star^2)^2} + \kappa_{\rm turb}^2 \frac{\tau_\star}{1+\tau_\star} = 1. \label{cond_turb}
\end{equation}

After some algebra, Eq. \ref{cond_turb} can be rewritten as a fifth order polynomial in $\tau_\star$:
\begin{align}
p(\tau_\star) & = (\kappa_{\rm l}^{2}+\kappa_{\rm turb}^{2}-1)\tau_\star^{5}+ (4\kappa_{\rm l}^{2}+2\kappa_{\rm turb}^{2}-2)\tau_{\star}^{3}\nonumber \\& + (\kappa_{\rm l}^{2}-1)\tau_{\star}^{4} +(4\kappa_{\rm l}^{2}-2)\tau_{\star}^{2}+(\kappa_{\rm turb}^{2}-1)\tau_{\star}-1   =0.\label{poly}
\end{align}
Unfortunately, there is no explicit expression for the roots of a fifth degree polynomial
\footnote{Solutions to third and forth order polynomials by radicals were known already in the 16th century. The first attempts of a proof of no analytic formula for the fifth degree was presented by Paolo Ruffini (\citeyear{ruffini1799}). His proof was incomplete and corrected by Niels Henrik Abel (\citeyear{abel1824}). This is known as the Abel-Ruffini theorem. Later, it was superseded by what is known today as Galois theory \citep{galois1846}, which was published postmortem only in 1846, 14 years after the tragic death of {\'E}variste Galois at 1832.}, but the roots can be found numerically. 

\begin{figure}
  \includegraphics[width=1.\linewidth, width=8.5cm]{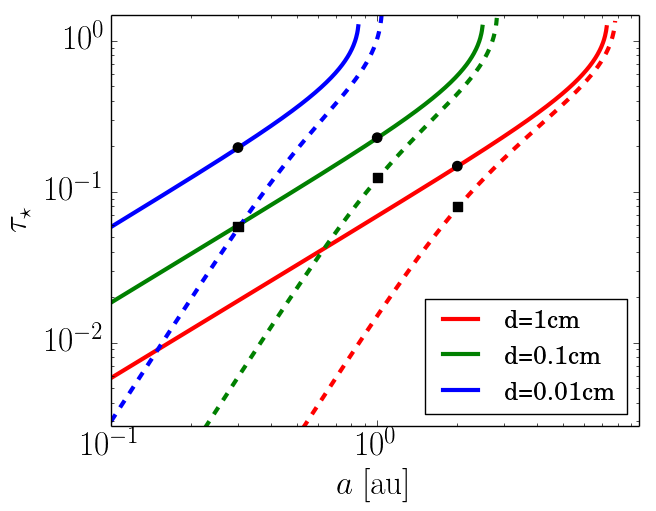}
\caption{Critical stokes number $\tau_\star$ as a function of the orbital separation $a$. Each line is the solution to Eq. \ref{poly}. Solid lines are solutions of the laminar velocity only ($\kappa_t\equiv0$ in Eq. \ref{poly}). Dashed lines are the solution with both laminar and turbulent velocity, with the $\alpha$-viscosity equals to $\alpha=0.01$. Red (top), green (middle) and blue (bottom) lines correspond to detaching grain sizes of $1,0.1, 0.01\ \rm cm$, respectively. Black circles indicate numerical integration of the erosion equation with laminar velocities for $a=0.3,1,2$ au and $d=0.01, 0.1, 1$ cm, respectively. Black squared indicate the same numerical integration, but with both laminar and turbulent velocities. } 
\label{fig:st_vs_a}
\end{figure}

Figure \ref{fig:st_vs_a} shows the critical Stokes number $\tau_s$ as a function of the orbital separation. Solid lines are the solution of Eq. \ref{tau_crit} with only laminar disk considered, while dashed lines are the solution to the turbulent disk, Eq. \ref{poly} with $\alpha=10^{-2}$. Eq. \ref{poly} had been solved numerically using the \texttt{numpy.polynomial} module. The solution is the smallest positive real solution. Each line represents different grain size that determines the threshold velocity in Eq. \ref{eq:vth}. Generally the critical Stokes number is a decreasing function on the radial separation. The larger the size of the pebbles, the farther in disk will erosion take place. 

\subsection{Disk Structure and Evolution}
The calculation of the critical Stokes number was done under the assumption of a Minimal Mass Solar Nebula \citep[MMSN]{hayashi81,PeretsMurray2011} background gas density. In reality, disk profiles could vary in shape and slope \citep{rc14}, and the gas density may vary due to various global and local effects. Transitional disks of depleted gas density are favourable for the formation of super-Earths \citep{lc2016}, and the formation of ice-giants requires the core to form relatively late in order to avoid runaway gas accretion \citep{bitsch2015}. 
SI was considered and found to be more efficient following disk evolution in depleted disks, where the metallicty is artificially enhanced \citep{carrera17}. 

Here we focus on the global disk dissipation and do not discuss local and/or transient effects, which could potentially be important, but are beyond the scope of the current study. We demonstrate the dependence on the results on the different gas densities. 

Observations of young clusters show that protoplanetary disks live only a few Myrs and could be fitted with exponential time dependence \citep{mamajek2009}. We assume for simplicity that the gas density follows an exponential decay law, $\rho_g(t)=\rho_g(0)\exp(-t/\tau_{\rm disk})$, where $\tau_{\rm disk}\approx 3\ \rm Myr$. Since $v_{\rm th}\propto \rho_g^{-1/2}$, the dimensionless parameter $\kappa_{\rm l}$ will decrease until the erosion will stop. For laminar velocity, the critical Stokes number depends on time via $\kappa_{\rm l}$, which will approach $\sqrt3/2$ at a finite erosion-stopping time 
\begin{equation}
    t_{\rm es}=\tau_{{\rm disk}} \ln\frac{4d\rho_{0}\eta^{2}v_{K}^{2}}{3A_N\gamma}.
\end{equation}
 At this time, the critical Stokes number will increase up to a limiting value of  $\tau_\star(\kappa_{\rm l}\to \sqrt3/2^{+})=\sqrt2\approx1.414$. For our fiducial values of $d=0.1\ \rm cm$ at $1\ \rm au$, the erosion-stopping time is $t_{\rm es}\approx 1.96 \tau_{\rm disk} \approx 5.9\ \rm Myr$. The result sensitively depends on the location in the disk. At larger radial locations $t_{\rm es}$ is reached faster since $\kappa_{\rm l}$ is smaller there, and vice versa. 
 
 For turbulent velocities, both $\kappa_{\rm l}$ and $\kappa_{\rm turb}$ will decrease with decreasing gas density. The critical Stokes number will increase, and generally larger Stokes numbers are possible. The erosion-stopping time is hard to compute analytically, but we expect it to be similar to the time obtained for the laminar case. 
 
 To summarize, the critical Stokes number increases with time as the disk is depleted. Therefore, assuming some supply rate $\dot{M}_{\rm supp}(t)$ of larger boulders and planeteismals (e.g. from pebbles drifting from larger separations where erosion was inefficient), the time-dependent erosion will leave traces of eroded material with time-dependent critical Stokes number $\tau_{\star}(t)$. The rate of erosion of larger boulders leading to production of grains/pebbles by the aeolian-erosion is $dN(\tau_\star)/dt = \dot{M}_{\rm supp}(t)/m_d(\tau_{\star}) \propto \tau_{\star}^{-3}$, where $m_d(\tau_{\star})$ is the mass of the grain at Stokes number $\tau_{\star}$. In principle, the production rate can be integrated to obtain the total number of new grains at a given time, but the integration is not trivial, since both $\tau_{\star}$ and $\dot{M}_{\rm supp}(t)$ could have complicated dependence on time. The number of new grains should decrease as the Stokes number increases.
 
\section{discussion and Implications} \label{s3}

\textbf{Size distributions:} The initial size distribution of disk solids is usually considered to be following a power-law with index $q$ ($n(r) \propto r^{-q}$). Observations of interstellar dust indicate that $q=3.3-3.6$ \citep{mathis77}. Evidence of multiple grain size populations had also been detected in molecular clouds  \citep{pagani10, and13} and in protoplanetary discs \citep{banzatti11, jin19}. The actual formation channels for boulders beyond the drift and fragmentation barriers are  debated. Various mechanisms have been suggested to overcome the growth barriers, such as local pressure maxima, particle pile-ups, rapid coagulation etc. (see sec. 4.3 in \citealp{armitagebook} and references therein.). Nevertheless, a large reservoir of $\tau_s \sim 1$ pebbbles is the starting point of the pebble accretion paradigm, and the numerical SI study of \cite{krapp2019} uses a wide range of sizes up to $\tau_s \sim 1$. The interstellar pebble and planetesimal reservoir could have been captured in most stages of the protoplanetry disk lifetime \citep{EvgeniHagaiSeeding}, or at an earlier stage during the molecular could phase \citep{pfalzner19}, which would enrich the protoplanetary nebula with an abundance of pebbles and noulders. We remain agnostic to the exact mechanism that forms these boulders and assume that a large reservoir exists, similarly to the standard pebble accretion scenario and other studies that assume an initial size-distribution (e.g. \citealp{krapp2019}) .

Regardless of the theoretical and observational uncertainties, the power law is expected to be steep. At $t=t_0$ the distribution is strictly a power law. As time progresses, dust will grow and the minimal size will increase. In addition, particles with $\tau_s>\tau_\star$ will be eroded to smaller pebbles with $\tau_\star$. If the growth is slow or inefficient, there will be little effect on the underlying distribution, since the total mass is dominated by the lighter dust particles. The only changes in the underlying power-law distribution are the boundaries of the minimal and maximal sizes, shaped by growth and erosion (and other barriers), respectively.

Nevertheless, the shaping of the dust size distribution could have effect on a \textit{local} scale. Since each radial separation $a$ determines a typical Stokes number $\tau_\star(a)$, different locations will have different typical dust sizes, which could in turn serve as ubiquitous mono-dispersed local population. This population can be important for the onset of other growth mechanisms as described below.

\textbf{Streaming Instability:}
Particles with $\tau_s > \tau_{\star}$ rapidly erode to $\tau_{\star}$ on dynamical timescales, much faster that the growth of SI. Thus, $\tau_{\star}$ is a natural upper limit for the allowed Stokes numbers for the initial multi-species size distribution. The inner parts of the disk will have lower $\tau_{\star}$. Although this natural upper limit is considered a barrier, it could actually help catalyze planet formation via SI.

Recently, \cite{krapp2019} have provided the first systematic study of the linear growth of the multi-species SI. They varied the minimal and maximal ranges of the Stokes number, the number of species $N$ and the local dust-to-gas density, $\epsilon$, and studied the timescale for the growth of the most unstable mode in each case. The most striking conclusion is that the convergence was not achieved with increasing number of species. In particular, even for favourable conditions with $\epsilon=1$, convergence was achieved for $\max(\tau_s)=0.1$ after $N\sim 100$ species, but for $\max(\tau_s)=1$ the timescale for the growth of the unstable mode is linearly increasing with the number of species, and does not seem to converge (see Figures 2 and 4 of \citealp{krapp2019}).

By truncating the maximal range of $\tau_s$ to $\tau_\star$, the SI mechanism can achieve convergence. Convergence is typically achieved for $\max{\tau_s} \lesssim 0.1$. Thus, the SI is favourable in areas in the disk for which $\tau_\star \lesssim 0.1$, which we find to be the regions inward to $\sim 1\ \rm au$, pending on dust size, disk model and turbulence levels. The boundaries of these areas, where $\tau_\star \approx 0.1$, could therefore be the most favourable areas for SI, since this is the optimal Stokes number at which SI is effective with the lowest possible metallicity $Z\approx 0.03$, as shown in \cite{yang17}. 

\textbf{Pebble accretion:} SI is a growth mechanism for the first planetesimals. Once planetary cores of $\gtrsim 10^{2-3}\ \rm km$ are formed, further growth is proceeded by accretion of pebbles until a critical core mass is reached, where runaway gas accretion begins leading to gas/ice giant formation. The efficiency of pebble accretion depends on their coupling to the gas, i.e. their Stokes number. Pebbles with $\tau_s \lesssim 10^{-3}$ are well coupled to the gas flow and unaffected by the core. Pebbles with $10^{-3}\le \tau_s \le 0.1$ are affected by core's gravity, but contribute less to the overall collisions and accretion rates \citep{lj2012}. Pebbles with $\tau_s \gtrsim 0.1$ are accreted onto the protoplanet when the impact parameter is within the Hill-sphere. Pebbles with $\tau_s \approx 1$ are attracted from wider distances, but the horseshoe orbits with small impact parameters are lost (see Fig. 7 of \citealp{lj2012}). The overall accretion rates are faster for Stokes numbers in the range of $0.1 \lesssim \tau_s \lesssim  1$, for large enough protoplanety core of  $10^3\ \rm km$ as seen e.g. in Fig. 10 of \cite{ok2010}.

The radial erosion-induced stratification of dust sizes plays a similar role in the efficiency of pebble accretion. Similarly to SI, there are favourable regions in the disk where the critical Stokes number is around $\tau_{\star} \approx 0.1-1$, where pebble accretion is most probable. Since these are generally regions close to $1\ \rm au$ and inwards, the accreting cores are unlikely to form gas giants. 
Only for boulders composed of relatively large dust grains of $d\approx 1\ \rm cm$, could erosion be effective up to larger distances of $\sim 7\ \rm au$, which is compatible with the formation locations of ice/giant planets. Evolved disks have lower densities, therefore less erosive, and even larger-grains composition (or closer separation) is required to be effective. 

\textbf{Caveats:} 
In the derivation of Equations (\ref{tau_crit}) and (\ref{poly}) we used dimensionless quantities. In reality, there is a limitations to the smallest Stokes number available. The Stokes number is defined as $\tau_s \equiv  (\pi/2) \rho_p d/\Sigma_g$. For our disk models, it is roughly $\tau_s \sim 10^{-3} (d/{\rm cm}) (a/\rm au)^{3/2}$. Thus, for size of $d=1\ \rm cm$ the minimal stokes is $\sim 10^{-3}$, which increases to $\sim 0.02$ at $a \approx 7\ \rm au$. Obviously, the erosion cannot grind-down boulders to sizes smaller than the fundamental composing-grain size, $d$, therefore there is a physical limitation on the minimal Stokes number in our formalism.

Growing boulders and plenetesimals can be porous and have various sizes and different densities and cohesive forces. From Eq. (\ref{eq:vth}) it is evident that detaching larger grains is easier than smaller ones. Thus, if an eroding object is composed of grains of various sizes, only grains above some threshold can be detached, which will affect the structure of the growing boulders and requires further study.

The erosion timescales are usually shorter than the radial drift times, but the drift itself is much faster than the disk's lifetime. As the particle will drift inward, its critical Stokes number will keep decreasing due to the decrease of the threshold velocity. Obviously, with no drift stopping mechanism, the body will be lost. Nevertheless, even if the body is lost, some of the fractions of the detached grains during the erosion process may survive and serve as reservoirs for the later growth mechanisms.  

It is also tempting to apply our formalism for large pebbles of sizes $\sim 10\ \rm cm$, since they are more favourable to efficient erosion. However, the aeolian-erosion formalism is relying on the assumption that the cohesive forces are linearly proportional to the dust grain size $d$. The proportionality constant was derived experimentally for small grain sizes of $\mu$-size. We extrapolated the linear behavior up to $\lesssim 0.1\ \rm cm$ pebbles in \cite{AeolianErosion1}, largely based on laboratory experiments of \citep{paraskov2006} for $0.05\ \rm cm$ size grains which seem to be consistent with our derived erosion rates. It is unclear if the linear proportionality could be extended beyond $1\ \rm cm$ scale. On the other hand, erosion of smaller grains from the surface may destabilize and weaken the cohesion of larger grains, possibly attached through contact with smaller grains. In this case erosion might be even more efficient. More generally, the nature of the forces that bring together the planetesimals which are composed of pebbles could be different and depend on the composition, porosity and equation of state, as well as self-gravity for the larger objects. We therefore caution using our model to larger dust/pebble sized and defer it to future studies.  

\section{summary} \label{s4}
In this letter we showcased that aeolian-erosion can efficiently grind-down solids in protoplanetary disks into smaller grains/pebbles down to the point where they are coupled to the gas flow. The strength of the coupling is measured by the critical Stokes number $\tau_{\star}$, (Eq. \ref{tau_crit}, \ref{poly}) which in turn depends on the ratio of the threshold velocity $v_{\rm th}$ (Eq. \ref{eq:vth}) and the typical relative laminar and turbulent velocities, and on the size of the detaching grains/pebbles $d$. The dependence can be related to the radial location on the disk (Fig. \ref{fig:st_vs_a}), and the general trend is that $\tau_{\star}$ is decreasing with decreasing radial location, until some critical separation where aeolian-erosion becomes inefficient.

Growth of planetesimals due to the streaming instability and the growth of planetary cores due to pebble accretion rely on large numbers of pebbles with 'optimal' Stokes numbers with non-trivial coupling with the gas. A wide size-distribution of small particles slows down the growth, since fewer particles participate, and complex coupling between different sizes may play a role and hinder the growth, therefore simplified assumptions in modelling of these processes through the use of ubiquitous, mono-sized pebbles is heavily criticized. However, as we show here, aeolian-erosion processes naturally produce particle sizes of typical Stokes number, depending on the radial separation. Erosion may therefore allow for a realistic, naturally produced limited pebble-size range. Optimal Stokes numbers are a natural consequence and are expected to then be present at preferred locations. The critical Stokes numbers depend not only on locations but also on time. Evaporating disks with lower gas density increase the critical Stokes number with time. Therefore, depleted disks (at later times or with local cavities) are better sites for planet/planetesimal formation mechanisms that require non-trivial coupling (e.g. $\tau_\star \approx 0.1-1$) of gas and dust, such as the streaming instability or pebble accretion.

\section*{Acknowledgements}
 
We thank Jake Simon, Allona Vazan, Andrew Youdin and Yanqin Wu for useful discussions. EG and MR acknowledge useful discussions at the Rocky Worlds workshop at the KICC, which initiated this work. HBP acknowledges support from the MINERVA center for "Life under extreme planetary conditions" and the Kingsley fellowship at Caltech.




\bibliographystyle{aasjournal}
\bibliography{references} 








\end{document}